\documentclass[traditabstract]{aa} 
\usepackage{graphicx}
\usepackage{natbib}
\bibpunct{(}{)}{;}{a}{}{,} 
\usepackage{txfonts}
\newcommand{\xmm}{{\small \it XMM-Newton}}
\newcommand{\chandra}{{\small \it Chandra}}
\newcommand{\ei}{EI~Eri}
\begin{document}
   \title{Variability of a stellar corona on a time scale of days}
\subtitle{Evidence for abundance fractionation in an emerging coronal active region}

   \author{R. Nordon \inst{1,2} \and E. Behar\inst{3} \and S. A. Drake\inst{4}
          }

   \institute{
	Max-Planck-Institut f\"ur Extraterrestrische Physik (MPE), Postfach 1312 85741 Garching Germany
         \and
   	School of Physics and Astronomy,
	    The Raymond and Beverly Sackler Faculty of Exact Sciences,
	    Tel-Aviv University, Tel-Aviv 69978, Israel.
	    \email{nordon@astro.tau.ac.il}
         \and
             Physics Department, Technion Israel Institute of Technology, Haifa 32000, Israel.
	     \email{behar@physics.technion.ac.il}
	 \and
	     NASA/GSFC, Greenbelt, Maryland 20771, USA. \email{stephen.a.drake@nasa.gov}
             }

   \date{Received...; accepted...}

 
  \abstract
{Elemental abundance effects in active coronae have eluded our understanding for almost three decades, since the discovery of the First Ionization Potential (FIP) effect on the sun. The goal of this paper is to monitor the same coronal structures over a time interval of six~days and resolve active regions on a stellar corona through rotational modulation.
We report on four iso-phase X-ray spectroscopic observations of the RS CVn binary \ei\ with \xmm , carried out approximately every two days, to match the rotation period of \ei .
We present an analysis of the thermal and chemical structure of the \ei\ corona as it evolves over the six days.
Although the corona is rather steady in its temperature distribution, the emission measure and FIP bias both vary and seem to be correlated.
An active region, predating the beginning of the campaign, repeatedly enters into our view at the same phase as it rotates from beyond the stellar limb.
As a result, the abundances tend slightly, but consistently, to increase for high FIP elements (an inverse FIP effect) with phase.
We estimate the abundance increase of high FIP elements in the active region to be of $\sim$75\% over the coronal mean.
This observed fractionation of elements in an active region on time scales of days provides circumstantial clues regarding the element enrichment mechanism of non-flaring stellar coronae.}


   \maketitle
%

\section{Introduction}

The best studied cool-star corona is that of our own sun, which also provides the only spatially resolved stellar coronal source of Extreme Ultraviolet (EUV) and X-rays. 
From X-ray images of the sun we know that the solar corona in its active state is non-uniform and that a significant fraction of the emission in these bands comes from magnetic loop structures concentrated in active regions around photospheric sun-spots. Evidently, these structures evolve in brightness on time scales of hours. The morphology as well as the metal abundances vary on longer time scales of days. 
Coronal structures on the Sun may survive for a few weeks, although tracking them continuously from the ground is difficult 
due to the $\sim 24.5$~day solar rotation period.

Active coronae of other cool stars can be up to five orders of magnitude more luminous and an order of magnitude hotter than the solar corona.
Although unresolved,  they show many characteristics similar to those of the solar corona, and are therefore believed to be structured and to evolve in a similar way.
For the most part, late-type stars have envelopes in which convection and differential rotation interact to drive a dynamo, which produces magnetic flux tubes. The flux tubes float into the upper atmosphere and eventually the corona. 
It is generally unclear, however, whether stellar coronae actually have larger covering of their stellar photosphere, or even qualitatively different morphologies, such as inter-binary magnetic structures.
Stellar coronae show high variability, often attributed to continuous micro-flaring, which is observed on the sun, but perhaps with different statistics \citep[e.g.,][]{Caramazza2007}.

Solar coronal chemical abundances are different from the solar photospheric composition.
The solar corona itself is non-uniform in that respect.
While in coronal holes, the abundances are close to photospheric, abundances in active regions and in the solar wind are significantly different.
The abundance patterns depend on the first ionization potential (FIP) \citep{Meyer1985, Feldman1992, Sheeley1995}. Elements with FIP$<$10~eV (low FIP) are enriched in the corona relative to elements with FIP$>$12~eV (high FIP). Averaging over the entire solar disk, the enrichment factor is about 4 \citep[][ and references therein]{Feldman2002}.
Such a FIP bias, in varying intensities, has been reported for many coronal structures, from closed magnetic loops to streamers \citep[see reviews by ][]{Kohl2006, Doschek2010}.
Currently there is no generally accepted model that explains the FIP bias. 

The FIP effect became even more puzzling when high resolution X-ray spectra from \xmm\ and \chandra\ revealed that active stellar coronae do not follow the solar FIP pattern. In some cases, the high-FIP elements are even enriched compared to the low-FIP ones, an effect labeled the inverse-FIP (IFIP) effect \citep{Brinkman2001}. 
Later studies revealed that FIP and IFIP biases are correlated with coronal activity (and age): 
Highly active stars show an IFIP effect, while less active coronae have a solar FIP bias \citep{Audard2003, Telleschi2005}, implying a transition on stellar-evolution time scales  of Gyrs.

On the face of it, the correlation with activity supported previous suggestions that abundances may be linked to flares 
\citep[e.g.,][]{Guedel1999, Osten2003, Guedel2004, Nordon2006}. However, direct systematic measurements of abundance variations during flares show that regardless of coronal FIP/IFIP bias, during large flares the abundances tend to photospheric values \citep{Nordon2007, Nordon2008}. The mechanism for FIP/IFIP is, therefore, related either to the long-term evolution of coronal structures, or to the continuous micro-flaring that produces the quiescence emission, but not to large flares.

In recent years, a model has been proposed in which abundance fractionation is driven by magneto hydrodynamics (MHD) waves \citep{Laming2004, Laming2009, Laming2012}. This `pondermotive' force is controlled by the energy density of MHD waves at the bases of the coronal loops. \citet{Wood2012} studied a sample of X-ray faint ($L_X<10^{29}$~erg~s$^{-1}$) dwarfs and found no correlation with X-ray luminosity or surface flux, but a good correlation with spectral type instead. According to this relation M~dwarfs are IFIP biased while early G~dwarfs are FIP biased. \citet{Wood2012} claim that this shift from FIP to IFIP can be explained by the inversion of the pondermotive force at certain conditions.
The latter study is limited to X-ray faint dwarfs and may not be applicable to the typical highly active binaries from other studies that are more than two order of magnitude more luminous in X-rays, emitted mostly by a giant or sub-giant star (RSCVn type systems).

Despite the assumed similarity between solar and stellar coronae, we have lacked a study of the evolution of a stellar corona on time scales of days, which is the relevant time for coronal loop evolution on the sun. The goal of the observations presented in this paper is to seek and measure the variability of temperatures and abundances in an active corona over several days. In order to avoid confusion between rotational variability, which is inevitable due to the rapid rotation typical of active stellar targets and the corona being composed of discrete active regions as on the sun, and genuine coronal-activity variability, the present observing campaign is designed to return to the target each time at exactly the same orbital phase.

\subsection{\ei }
In order to carry out a few iso-phase observations, we chose the target of EI Eridani (\ei , HD 26337), an X-ray bright ($L_x = 10^{31.1}$~erg/s) RS CVn binary, whose orbital period is 1.95 days (168.5~ks), which is very similar to the exactly 2.0 day orbital period of \xmm\ around Earth.
This enabled \xmm\ to observe the exact same orbital phase of \ei\ during four consecutive orbits, and for sufficiently long exposures each time ($\sim35$ ks).
\ei\ is a nearby \citep[50.8~pc,][]{van Leeuwen2007}, rapidly rotating ($v\sin i = 51$km\,s$^{-1}$), non-eclipsing, rotationally locked, close RS~CVn binary with an inclination angle to the rotation axis of $\sim 56 ^\circ  \pm 4.5^\circ$ \citep{Washuettl2009a}.
These properties make \ei\ a popular target for Doppler imaging and starspot monitoring that indicate the primary star to be highly active.
\ei\ consists of a solar-mass G5IV primary and a much fainter M4-5 secondary, with a possible 19-year orbit of a third body \citep{Washuettl2009a}.

Long term photometric monitoring of \ei\ seemed to indicate a $\sim$11~yr brightness cycle that was conjectured to be an analog of the solar activity cycle \citep{Olah2002}, but unlike the sun,  long term Doppler imaging of \ei\ by \citet{Washuettl2009b} shows little variability of its cool starspots over 11 yrs. 
As for the hot corona, \citet{Osten2002} observed \ei\ with ASCA (and EUVE) and fitted its spectrum with a 2T model of $kT$~= 0.72 keV and 2.2 keV, but the abundances could not be tightly constrained with ASCA.
The \xmm\ CCD spectra of a short flare on \ei\ on 2009-08-06 during the present campaign was analyzed by \citet{Pandey2012}, who suggested the flare originated from a loop the size of $\sim R_*$ (primary), which is about a tenth of the binary separation distance. 
 
\section{Observations, Data, and Analysis}
\label{sec:obs_and_data}

\begin{table*}[t]
\caption{Observing log}
\centering
 \begin{tabular}{lllc}
  Obs. ID & Start time & End time & Exposure [ks]\\
  \hline \hline
  060317201 &2009-08-01T00:02:50 &2009-08-01T09:13:05 & 32.84\\
  060317301 &2009-08-02T23:06:16 &2009-08-03T09:24:46 & 36.92\\
  060317401 &2009-08-04T22:20:49 &2009-08-05T07:42:41 & 33.54\\
  060317501 &2009-08-06T20:59:39 &2009-08-07T07:31:30 & 37.48\\
 \end{tabular}
\label{tab:Observations}
\end{table*}

In Table~1 we list the log of the four \xmm\ observations of \ei .
Note the start times for each observation, which are separated by approximately the \ei\ orbital period of 168.5~ks.

\subsection{Data Reduction}
\xmm\ data were reduced using the Science Analysis Software (SAS - ver. 10.0.0) in the standard processing chains as described in the ABC Guide to \xmm\ Data Analysis\footnote{See http://heasarc.gsfc.nasa.gov/docs/xmm/abc},
and using the most updated calibration files (date 22.06.2010).
Pileup was checked using the {\it epatplot} routine of the SAS software package. No significant pileup was detected in either of the EPIC instruments and thus events from single to quadruple were retained.
Background was low during all the observations and no significant amount of time was lost.

\subsection{Spectral Analysis}

We measure the temperature and abundance properties of the \ei\ corona by fitting thermal plasma models to the observed spectra.
\xmm\ comprises several different instruments.  The longer wavelengths (6-38~$\AA$) are covered by the two high resolution reflection grating spectrometers (RGS) that resolve the emission lines. The high energy end up to 15~keV is covered by the two EPIC-MOS and single EPIC-pn CCDs, that provide lower spectral resolution.

In order to study the thermal and chemical composition of the source plasma, it is imperative to first properly account for its Emission Measure Distribution that is defined as

\begin{equation}
EMD(T) = n_e n_H dV/dT.
\label{eq:EMD_definition}
\end{equation} 

\noindent where $n_e$ and $n_H$ are the electron and proton number densities, respectively, and $dV/dT$ is the temperature distribution of the plasma volume.
The $EMD$ is usually characterized by several discrete temperature bins.  Here, we use seven temperature bins with identical abundances.
To allow for a uniform analysis of the four observations, the temperatures of the middle five components $kT=$ 0.7, 1.0, 1.5, 2.0, and 2.5 keV are held fixed, while the lowest and highest temperatures are allowed to vary and are determined by fitting the model to the observed spectra.
We fit the observed spectra using the XSPEC \citep{Arnaud1996} software package and a VAPEC \citep{APEC01} plasma spectral model.
For high statistics, we fit the two RGS, EPIC-pn, and EPIC-MOS1 spectra of each observation (orbit)  simultaneously.
EPIC-MOS2 was used in timing mode and its spectra are systematically offset by a small amount from EPIC-pn and EPIC-MOS1 that agree with each other very well. For this reason, we exclude it from the spectral fitting.
We ignore the very low energy ($E< 0.35$ keV) portions of the CCD spectra.
Visual inspection of each RGS spectrum indicates the models do an overall good job of fitting the lines.  
See the appendix for the spectra and fitted model. 
The absorption column density towards \ei\ is negligible \citep[$N_H=1.2\times10^{20}$~cm$^{-2}$, ][]{Pandey2012} and was not included in the model.

The seven plasma components provide only a general idea of the temperature distribution in the plasma.
It is known \citep[see most recently, ][]{Testa2012} that the $EMD$ reconstruction should be treated with caution.
More specifically, the $EMD$ is highly degenerate between adjacent components.
Therefore, we keep track of the covariance matrices (between parameters), which are fully taken into account when propagating the errors of all derived quantities.
Due to this degeneracy, the more important and reliable quantity is the cumulative, or progressively Integrated Emission Measure $\left( \int EMD\, dT \right)$, which we denote by $IEM$.
The calculated errors for the $IEM$ are more meaningful than those of the $EMD$, as they are essentially independent of the choice of temperature bins, as opposed to errors on the $EMD$ that are affected by the arbitrary choice of temperatures and by the proximity of the other temperature components.
Monte Carlo (MC) simulations are used for error estimates.
The MC simulation randomly rolls a new set of $EMD$ values according to the correlation matrix and individual-component 
errors obtained from the best-fit solution. 
The central 68\% of the distribution for each point in the $IEM$ is taken as the equivalent of the 1-$\sigma$ confidence interval.
For a more detailed description of the error propagation see Appendix~\ref{app:MC_simulation}.

\section{Results}
\label{sec:days_scale_evol}

\subsection{Light Curves}

\begin{figure}[t]
 \centering
 \includegraphics[width=\columnwidth]{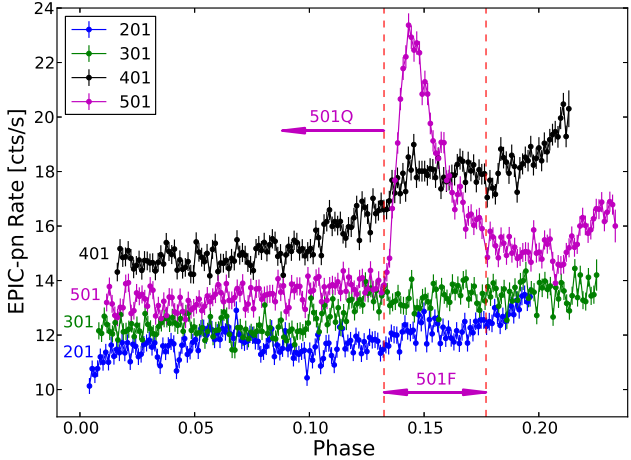}
 \caption{EPIC-pn light curves during the four consecutive \ei\ rotation orbits. The time axis has been folded to represent the rotation phase within the 1.95 days period of \ei .} 
 \label{fg:pn_lc}
\end{figure}

By synchronizing the \xmm\ exposures with the rotation phase of \ei , we are able to observe the same stellar surface during each visit, and to thus remove most of the rotation effects.
The EPIC-pn light curves of the four exposures are plotted in Figure~\ref{fg:pn_lc} versus the rotaional phase. 
Only about 20\% of the orbit, i.e., 35~ks or $\sim$0.4~days, is covered during each visit, and consecutive visits are separated by the remainder of the \ei\ period ($\sim$2 days).

Between phases 0 -- 0.1, all 4 exposures show a steady count rate with some variability on time scales of minutes, but at a low amplitude that is not much above the Poisson noise. 
This variability is typical of active coronae, and has been ascribed to micro-flaring.
The second half of each observation, i.e., phases 0.1 -- 0.2, features a gradual rise in flux that is interpreted as the recurrent appearance of an active region on that part of the star. 

This brightening with phase is most dramatic during the third observation (401), $\sim$4 days into the monitoring, but should still not be confused with a coronal flare that would tend to sharply rise within a few ks and quickly decay over $\sim 10$~ks.
Indeed, during the last exposure (obs. 501), a large flare is superimposed on the gradually increasing light curve, rising sharply ($\sim$1.5 ks) about half way through the exposure and decaying exponentially on a time scale of $\sim$4.8~ks.
For analysis purposes, we therefore further split observation 501 into its quiescence and flare periods, as indicated in Fig.~\ref{fg:pn_lc}, and refer to them as obs. 501Q and 501F, respectively.

The mean count rate between visits ($\sim 130$~ks) also varies, increasing within the first 4 days, between obs. 201 and 401 by $\sim 20\%$, and then decreasing by $\sim$30\% by the beginning of obs. 501. Interestingly, just before the flare and $\sim$5 days into the monitoring, the steady quiescent level is observed to be back to its initial level when the monitoring started (obs. 201, 301).

\subsection{Temperature}

\begin{figure}[t]
 \centering
 \includegraphics[width=\columnwidth]{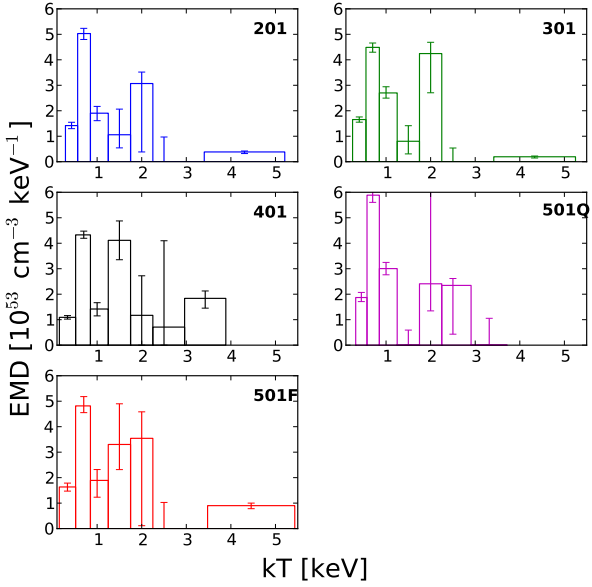}
 \caption{Emission measure distribution as obtained by the best-fitted 7T models of each observation.}
 \label{fg:EMD}
\end{figure}

\begin{figure}[t]
 \centering
 \includegraphics[width=\columnwidth]{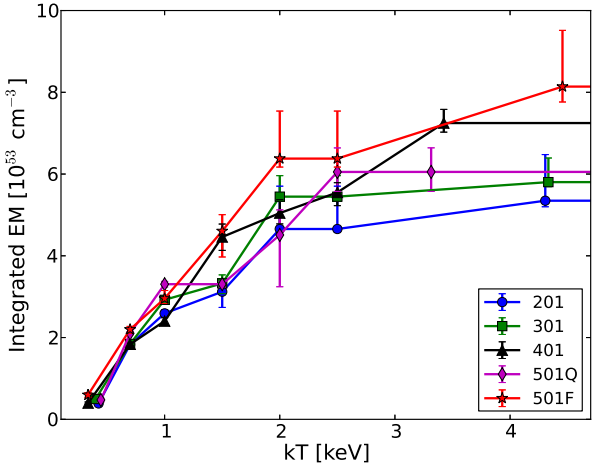}
 \caption{Cumulatively integrated emission measure distribution $IEM$ calculated from the $EMD$s of Fig.~\ref{fg:EMD}. Errors take into account the degeneracy between the different temperature components of the models.}
 \label{fg:IEM}
\end{figure}

In Figure~\ref{fg:EMD} we plot the $EMD$($T$) for each observation based on our 7-temperatures best-fitted models, where each $EMD$ bin is due to a single temperature component. 
The width of each plotted bin is arbitrary and designed to give an impression of a continuous distribution, although in reality these are iso-$T$ components.
As with the light curves, the $EMD$ of observations 201, 301, and 501Q are fairly similar, primarily showing emission of plasma up to $kT \simeq 2$~keV, while the $EMD$ of observation 401 features some excess high-$T$ emission up to $kT \simeq 4$~keV.
During the flare (obs. 501F), significant emission measure is observed at $kT \geq 4$~keV.

The $IEM$s for each observation are then plotted in Fig.~\ref{fg:IEM}.
The $IEM$ in observations 201, 301 and 501Q are consistent with no variation in the total emission measure (see the high temperature end of Fig.~\ref{fg:IEM}), and show very small variations below 1~keV.
On the other hand, the \ei\ corona during observation 401 has less emission measure below 1~keV, but a higher total, implying a significant excess of plasma at higher temperatures, mostly between 2.5--3.5~keV.
During the short flare (obs. 501F), one can see an excess of emission measure between 1--2~keV and another small component above 3~keV that brings the total emission measure to a level $\sim 40\%$ higher than the quiescent states of obs. 201, 301, and the pre-flare 501Q, which is also somewhat lower than during obs. 401.
In short, the emission measure of the quiescent corona of \ei\ can vary between rotation orbits, and on time scale of days, but mostly in the 1 --2 keV temperature regime.
Excess emission measure of hotter plasma up to 4 -- 5 keV is observed both in active regions (obs. 401), and more dramatically during the flare (obs. 501F).

\begin{figure}[t]
 \centering
 \includegraphics[width=\columnwidth]{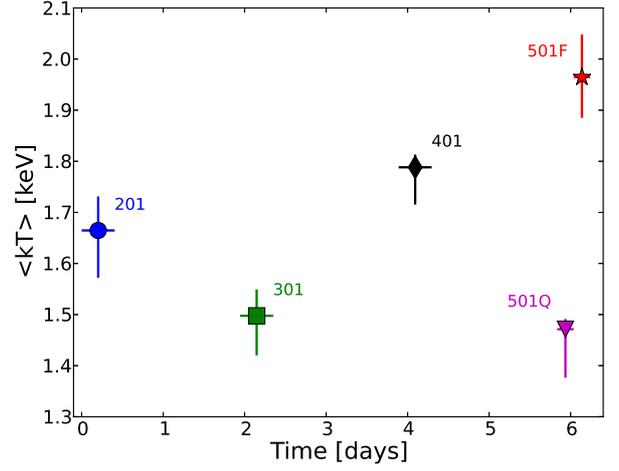}
 \caption{Variability of the $EMD$-weighted  mean temperature. Horizontal error bars indicate the exposure time of each observation.}
 \label{fg:kT_effective}
\end{figure}

A concise way to quantify the plasma conditions is through its mean temperature $<kT>$, weighted by the $EMD$ of each component.
The errors are again computed using an MC simulation, while taking into account the correlations between the fitted $EMD$s of the various model components.
The resulting $<kT>$ values are plotted in Figure~\ref{fg:kT_effective} for each observation and against time elapsed from the beginning of the first observation.
A similar pattern as before emerges.
When the source is brighter (e.g., Obs. 401) it is also hotter on average.
Obs. 201 and 301 do not strictly follow this pattern, but the temperature decrease from Obs. 301 to Obs. 201 is insignificant.
In terms of $<kT>$, we observe temperature variations on time scales of days, and most dramatically before and during the flare. Temperature variability during quiescence is more subtle, with a mean $kT$ of 1.60~keV and a standard deviation of  0.13~keV.  

\subsection{Abundances}
\label{sec:abundances}

\begin{figure}[t]
 \centering
 \includegraphics[width=\columnwidth]{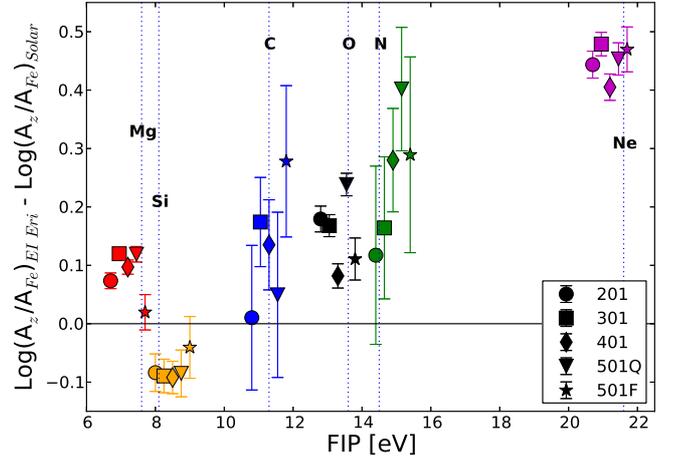}
 \caption{Coronal abundances relative to Fe and expressed relative to the solar X/Fe abundance ratio \citep{Asplund2009}. Points from different observations are progressively shifted to the right along the horizontal axis in the order in which they were taken ($\sim$2 days apart) to enhance visibility and to emphasize the time variability.}
 \label{fg:abundances_Fe}
\end{figure}

\begin{figure}[t]
 \centering
 \includegraphics[width=\columnwidth]{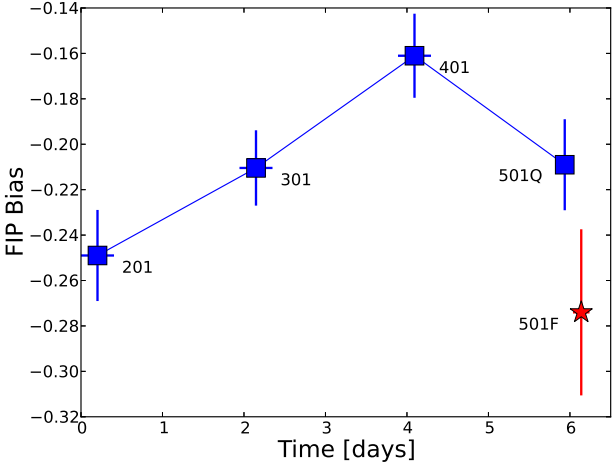}
 \caption{FIP bias ($FB$ from Eq.~\ref{eq:FB}) for each observation and as a function of time. 
Horizontal error bars indicate the exposure time of each observation.
 }
 \label{fg:FIP_bias}
\end{figure}

In Figure~\ref{fg:abundances_Fe} we plot the relative abundances of the major elements with respect to Fe, in solar-abundance units \citep{Asplund2009}, and as a function of their first ionization potential (FIP).
In the plot, the data points are progressively shifted along the x-axis according to the order in time in which they were measured,
thus creating a time-like sequence for each element.
The abundances are expressed relative to the Fe abundance instead of the more common absolute values relative to H.
The absolute Fe abundance with respect to H is 0.32, 0.32, 0.38 \& 0.31 solar in observations 201, 301, 401 \& 501Q respectively, with a typical error of $\pm$0.03.
In the APEC model and the global spectral fit approach in the XSPEC software that we use, the absolute abundance is constrained by the flux in the Fe lines and the overall fit to the continuum. While lines are emitted from a limited range of temperautes, continuum flux at all RGS wavelengths is contributed from all temperature components.
Absolute abundance is therefore sensitive to the underlying temperature distribution, which is not so well constrained at the high temperatures.
Conversely, estimates of abundances relative to Fe rely on well-measured line ratios that tend to emerge from a limited and overlapping range of temperatures.
The large number of observed Fe ions and lines, whose peak emissivities span a wide temperature range from $kT \sim 0.3 - 1.5$~keV makes these measurements significantly less sensitive to the temperature distribution in the plasma.
The selection of the reference element for describing abundances does not affect the general pattern of abundance variations.
We normalize the abundance ratios to those of the solar photospheric composition of \citet{Asplund2009} in order to plot them on the same scale and to discuss the FIP effect.

As seen in Fig.~\ref{fg:abundances_Fe}, \ei\ shows the typical inverse FIP effect observed in the most active coronae \citep{Brinkman2001, Audard2003}.
In particular, the Ne abundance is enriched by a factor $\sim$3 compared to Fe, and a lesser enrichment is observed for
C, N, and O (all high FIP) over Fe (low FIP),  
The abundance ratio of the low-FIP elements Mg and Si to Fe is approximately solar, although with respect to H, they are all sub-solar ($\sim 0.3$). 

Little variation of the abundances with the level of activity is observed.
The O (high FIP) abundance does seem to be low during the high-states of obs. 401 and 501F (flare), and high during the low-activity state of obs. 501Q.
This is reminiscent of the chromospheric evaporation effect of solar-composition material during flares in an otherwise (quiescent) inverse-FIP coronae, which was observed by \citet{Nordon2007, Nordon2008} in a sample of flares.
The highly enriched Ne abundance is slightly reduced in the high state of obs. 401.
However, despite the good signal, no significant variation in the Ne abundance can be detected during any other period, not even during the flare, which would be expected if chromospheric evaporation takes place. 
The Mg/Fe abundance ratio comes down by $\sim 20\%$ during the flare, but this can not be a FIP related effect, as both elements are low-FIP.

In order to better quantify the abundance variations, a single number that quantifies the FIP bias can be defined \citep[following][]{Nordon2008}

\begin{equation}
  FB = \left< \log(A_Z/A_{Z_\odot})_{\rm FIP<10\,eV} \right> - \left< \log(A_Z/A_{Z_\odot})_{\rm FIP>10\,eV} \right>
\label{eq:FB}
\end{equation}

\noindent where $\left< \log(A_Z/A_{Z_\odot}) \right>$ is the error-weighted mean abundance (in log) of the low/high FIP elements, relative to a set of reference abundances, in this case the solar abundances. 
We use the traditional solar distinction between low FIP ($<$10~eV) and high FIP ($>$10~eV).
A positive $FB$ value indicates a solar-like FIP bias, while a negative $FB$ value indicates an IFIP bias.

The FIP biases in the five observations (201, 301, 401, 501Q, 501F) are shown in Fig.~\ref{fg:FIP_bias}.
The $FB$ variations over the course of the present campaign are not dramatic, but still some trends can be identified.
The $FB$ in Fig.~\ref{fg:FIP_bias} generally varies with activity, namely the $FB$ increases, i.e., tends to (solar) photospheric values, as the activity increases, rising from obs. 201 to obs. 301, and further in obs. 401, and then decreasing in obs. 501Q when the count rate returns to its lower level. For comparison, see the light curves in Fig.\ref{fg:pn_lc}.
This trend breaks down during the flare when the activity strongly increases, but the $FB$ decreases.
At face value, this means that as the (quiescent) coronal activity increases, perhaps more small active regions appear, and more high-FIP elements  are preferentially brought up into the corona. 
This process is reminiscent of the evaporation of chromospheric material observed in the largest flares  on the most active (strongest IFIP) coronae \citep{Nordon2007, Nordon2008}.
On the other hand, the present flare on \ei\ does not follow this trend, but the weak abundance variation observed is consistent with the abundance variations observed in flares on coronae with similar quiescent $FB$ values of about $-0.2$ \citep[see Fig.~5 in][]{Nordon2008}.

\section{Rotation Effects}
\label{sec:Rotation_Effects}

The four exposures that all cover the same rotation phase allow us to study the dependence of temperatures and abundances on phase, by further breaking up each observation into smaller segments and co-adding the similar phase segments from each observation.
If active coronal regions are compact, and sparse in longitude,
then the rotation would create variations in the observed X-ray emission properties, as active regions rotate in and out of view, over the limb of the star.
If the active regions last longer than the monitoring time of 6 days (as solar active regions do), then stacking data from the same phase, but from different orbits, can increase the signal of this rotational modulation. 

During each full observation of $\sim$36~ks (Table~1), \ei\ rotates by 72 degrees, i.e., 20\% of the orbit.
The inclination angle of the rotation axis to the line of sight of $56 ^\circ$ implies only about 10\% of its surface gradually appears during the observation and a different 10\% gradually moves out of sight. 
We divide each observation into four quarters of $\sim 9$~ks each, during each of which the \ei\ system rotates by 18 degrees.
The data are then reduced in quarters and the spectra of each instrument are combined per phase bin.
The two late phase bins of Obs. 501 are excluded from the analysis, due to the flare (see Fig.~\ref{fg:pn_lc}).
Subsequently, we perform the same EM and abundance analysis as before, but for each phase-stacked spectra.

\begin{figure}[t]
 \centering
 \includegraphics[width=\columnwidth]{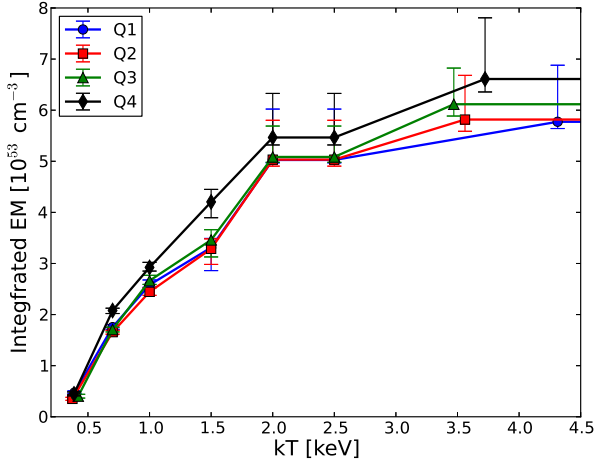}
 \caption{Cumulatively integrated emission measure distribution ($IEM$) as a function of the four phase bins Q1 - Q4.
}. 
 \label{fg:quarters_IEM}
\end{figure}

Figure~\ref{fg:quarters_IEM} shows the $IEM$ for each phase bin, Q1 -- Q4.
The first three phase bins show the same temperature structure. 
In the last phase bin (Q4), there is an excess of emission at temperatures below $\sim$1.5~keV,
which slightly increases the total $IEM$ of Q4, although it is just consistent with the previous quarters to within the errors.
This excess emission measure is a result of the rise in flux observed in the light curves (Figure~\ref{fg:pn_lc}) during the last quarter of all of the observations, but most conspicuously in Obs. 401.
A similar rise in flux can be seen also in the last quarter of Obs. 501, which is somewhat masked by the flare, and therefore excluded from the analysis.
Figure~\ref{fg:quarters_IEM} can be explained by an active region that is not visible at all during the two earlier quarters, that comes into view from beyond the limb during Q3 or Q4 (i.e., after $\sim$ 40 degrees of rotation), and that gradually brightens during the four days after the beginning of the \xmm\ monitoring, i.e., it becomes particularly bright during Obs. 401 and Obs. 501.

\begin{figure}[t]
 \centering
 \includegraphics[width=\columnwidth]{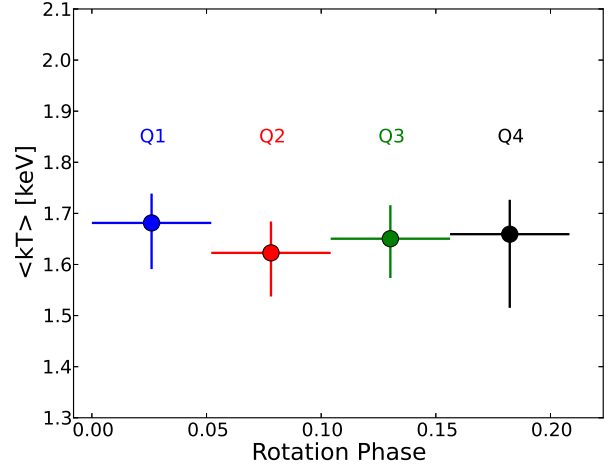}
 \caption{Mean temperature for the four rotation phase bins. Very little evolution is observed compared to the days-scale temperature variations (Figure~\ref{fg:kT_effective}).}
 \label{fg:quarters_effective_kT}
\end{figure}

In Figure~\ref{fg:quarters_effective_kT} we plot the mean temperature as a function of phase.
The value of $kT_\mathrm{eff}$ is very stable between the observed rotation phases, despite the added emission measure.
In fact, it shows much less scatter than the scatter from one observation to the other over several days (Figure~\ref{fg:kT_effective}).
The lack of temeprature variation is somewhat expected, since each phase represent a time-average of 4-6 days, and variations on shorter time scales tend to average out.
Note that here too, the two last quarters of Obs. 501 are excluded from the analysis due to the flare.
The emerging active region (presumably a magnetic loop) reaches the mean temperature of  ($kT \sim 1.5$~keV), which is very similar to the mean temperature of the coronal regions on the rest of the \ei\ observed area.

\begin{figure}[t]
 \centering
 \includegraphics[width=\columnwidth]{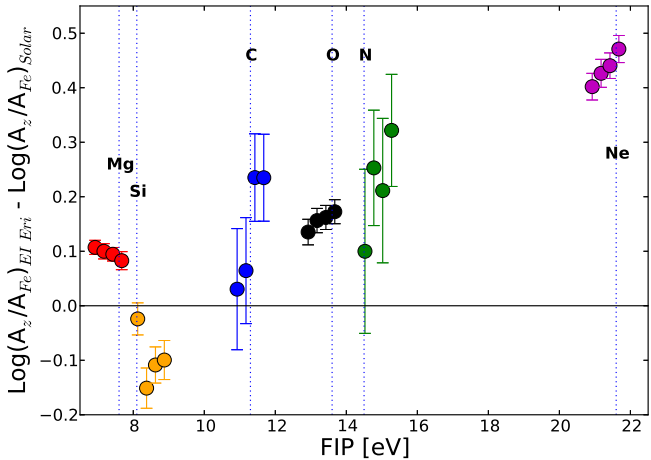}
 \caption{Coronal abundances relative to Fe and expressed relative to the solar X/Fe abundance ratio \citep{Asplund2009}. Points from different phase bins Q1 - Q4 are progressively shifted to the right along the horizontal axis to enhance visibility and to emphasize the variability with phase.
 }
 \label{fg:quarters_abundances_Fe}
\end{figure}

The mean abundances for each rotation phase bin are plotted in Figure~\ref{fg:quarters_abundances_Fe}. 
Again, 
the data points are progressively shifted along the horizontal axis, in order to give an idea of the evolution with phase.
The variation in abundance with phase for each element is very small, and statistically insignificant between consecutive bins.
However, Mg, O and Ne, the three elements with the best signal do appear to vary monotonously with phase.
The abundance ratio of Mg/Fe decreases while that of O/Fe and Ne/Fe increases.
The other elements tend to show a similar monotonic variation, although with much larger uncertainties.
While the trend is of low signal in each element, the repeated pattern in several elements makes it more believable, and less likely due to noise.
Moreover, since the trend of the low-FIP Mg is opposite to that of the high-FIP Ne and O, we suspect it is due to a FIP related effect and not due to only Fe changing with phase.

\begin{figure}[t]
 \centering
 \includegraphics[width=\columnwidth]{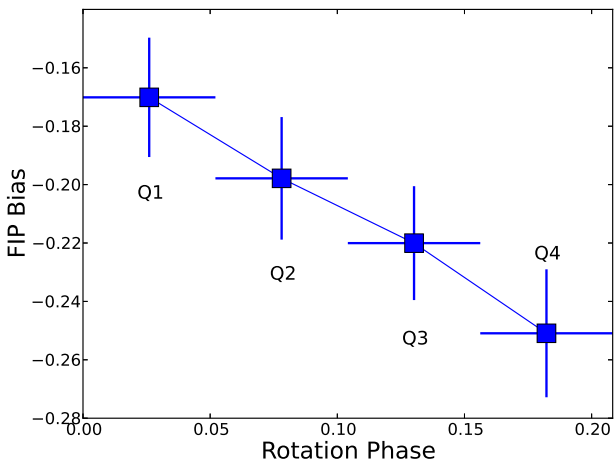}
 \caption{FIP bias ($FB$) as a function of rotation phase averaged over all available observations in four iso-phase intervals.  Horizontal error bars reflect the $\sim $9~ks duration in each bin. }
 \label{fg:quarters_FIP_bias}
\end{figure}

As in Sec.~\ref{sec:abundances}, we wish to enhance the individual abundance signals as to reflect potential FIP related effects by computing the FIP bias from Eq.~\ref{eq:FB} for each phase-stacked spectra.  The FIP bias as a function of phase is plotted in Fig.~\ref{fg:quarters_FIP_bias}. The $FB$ shows a mild, yet monotonous decrease with phase.
This can be interpreted as the emerging active region contributing an inverse FIP bias.
In other words, the active region (Q3, Q4) has slightly more high-FIP elements than the background corona (Q1, Q2).
Note that this time-averaged rotation effect is different from the time-evolution observed over days (Fig. 6), in which the FB increases with increased activity.
Here, the increased activity coming from a specific region on the star is the likely cause of the decrease in $FB$.

The weakness of the observed signal is not unexpected, as the fraction of the \ei\ surface rotating in and out of our view is not very large, $\sim 5\%$ between quarters.
Say we measure during rotation phase Q4, in which the active region is visible, a mean abundance $A^{(4)}_Z$ which deviates from the mean abundance during phase Q1 $A^{(1)}_Z$ (in which the active region is not visible) by $\alpha \approx 0.1$, i.e. $A^{(4)}_Z = (1+\alpha)A^{(1)}_Z$.
The total count rate is increasing with phase, indicating that this effect is likely due to a different abundance in the active region which is rotating into view, for which $A^{region}_{Z} = (1+\beta)A^{(1)}_Z$.
Since the contributions to the spectrum of both the background and the active-region abundances are weighted by their respective emission measures, and we attribute all the excess emission measure during Q4 over Q1 to the active region, then one can easily show that:
\begin{equation}
 \beta = \frac{\alpha}{1-(EM^{(1)}/EM^{(4)})}
\end{equation}
where $EM^{(4)}$ and $EM^{(1)}$ are the total emission measures in Q4 and Q1 respectively.
According to Figure \ref{fg:quarters_IEM}, the $EM$ ratio is $EM^{\rm (4)}/EM^\mathrm{1} \approx\ 1.15$. 
Therefore, the actual abundance variation in the active region by comparison with the spatial-mean background corona is at the $\beta \approx 0.75$ level, or a 75\% stronger IFIP effect than the mean corona.
Admittedly, this conclusion for \ei\ is based on a marginal signal and is, thus, associated with appreciable uncertainty, but it does point to significant abundance variations in a compact active region with respect to the mean corona.

\section{Discussion and Conclusions}

The light curves and abundance patterns in the corona of \ei\ show an active region appearing over the limb as the corona rotates with the star. 
Although on time scales of days, the increase of the activity level seems to lead to an increase of the FB (i.e., less IFIP bias, abundances more akin to the photosphere), the appearance of the active region with phase features an increase in high-FIP elements and a lowered FB (i.e. a tendency to a stronger IFIP bias).
Two different effects are at play: one is the evolution of the emission measure and coronal abundances on a time scale of days between exposures, spatially integrated and averaged over the entire visible surface of the star.
The other is an already-existing, compact and bright active-region that persists over the six days duration of the monitoring campaign and rotates into view towards the end of each exposure.
Thus, the active regions is manifested as variations with phase instead of time.

One cannot tell whether the days-scale variations in count rates, emission measure, and abundances are due to the evolution of existing coronal loops, or due to the appearance of new coronal loops and disappearance of old loops.
However, the correlation in which an increase in emission measure and count rates corresponds to a decrease in the IFIP bias (increase in FB, closer to photospheric composition) offers an interesting interpretation.
If the brightening is due to the emergence of new loops, and those appear initially with nearly photospheric composition, then the mean coronal abundances will tend towards photospheric composition during these times.
Conversely, during times in which no new loops emerge and existing loops age, evolve in composition and perhaps some fade, the mean abundances will tend to resemble the composition of the older, evolved loops, i.e. different than photospheric.

Indeed, on the sun, it has been reported that solar coronal loops evolve on time scales of days \citep[e.g.,][]{Widing2001}. 
Newly emerging structures tend to have photospheric composition, but develop the solar FIP bias within a few days. 
Moreover, on the sun, the typical enhancement of low FIP abundances relative to photospheric values by a factor of 4 is reached within 2-3 days and continues to evolve further, reaching a factor of 8 or more.
The analogous interpretation and one in agreement with the present observations is that new coronal loops on \ei\ emerge with photospheric composition and develop an IFIP bias over time scales of days.
The spatial mean IFIP bias is determined by mean age of the loops.
Thus, periods of gradual brightening of the corona are characterized by the emergence of young loops that reduce the global IFIP bias, while periods of gradual dimming are characterized by aging of existing loops that increase the global IFIP bias.
Note that the negative correlation between emission measure and IFIP bias as described above relates to the days scale evolution of a given stellar corona and is not at odds with the general positive correlation found between activity level and tendency to IFIP bias in stellar populations.

Rotational modulation in X-rays and EUV has been reported for HR\,1099 \citep{Agrawal1988, Drake1994, Audard2001}, from light curves, but phase dependent spectroscopy was not possible, due to the constant flaring of that source, which makes it difficult to disentangle between rotational modulation of active regions and their intrinsic activity variability.
A different, yet notable, phase dependent measurement is that of line profiles by \citet{Brickhouse2001}, which enabled latitude determination of active coronal regions on the stellar binary 44i Boo. 
The present measurement, however, is the first to measure phase dependent abundances.

The active region that appears towards the end of each exposure persists throughout the six days of the observing campaign. 
It therefore likely contains loops that are at least a few days old. In the scenario which we proposed above, these loops will have already evolved (over days, possibly starting even before the beginning of the campaign) and their abundances developed a strong IFIP bias, stronger than the coronal mean bias that includes younger loops.
Thus, as this region comes into view, the spatially averaged abundances tend towards a stronger IFIP bias (Figure~\ref{fg:quarters_FIP_bias}).

After showing that large flares tend to counter the coronal FIP effect \citep{Nordon2007, Nordon2008} - the present work promotes the notion that the abundances in individual coronal active regions, differ from the mean coronal abundances as well.
It is not clear whether large flares show coronal abundances that are closer to photospheric composition than the quiescent corona because of added material from the photosphere/chromosphere (chromospheric evaporation) or because these flares happen preferentially in young active regions where the abundances are closer to photospheric than the global mean.
The abundances measured during the flare observed in this campaign show an increase IFIP bias (more negative FB, Figure~\ref{fg:FIP_bias}) instead of the typical tendency to bring abundances towards photospheric composition.
If the flare is related to the limb active region that is strongly IFIP biased, it is possible that the mean abundances, weighted between the flaring active region, evaporated chromospheric material and surrounding quiescent corona end up being dominated by the abundances of the active region.

While the proposed scenario in which young loops emerge with photospheric-like composition and then develop over time-scale of days to an IFIP bias can explain our observations, it does not explain why the abundances evolve an IFIP effect instead of a solar-like FIP effect.
For a theory that attempts to resolve this issue see \citet{Laming2004, Laming2009, Laming2012}.

\begin{acknowledgements}
This work was supported by The Israel Science Foundation (grant \#1163/10) and by the Ministry of Science and Technology.
\end{acknowledgements}

\bibliographystyle{aa}


\newpage
\begin{appendix}

 \section{Spectra}
 \label{app:XMM_spectra}
 
\begin{figure*}[h]
 \centering
 \includegraphics[width=\textwidth]{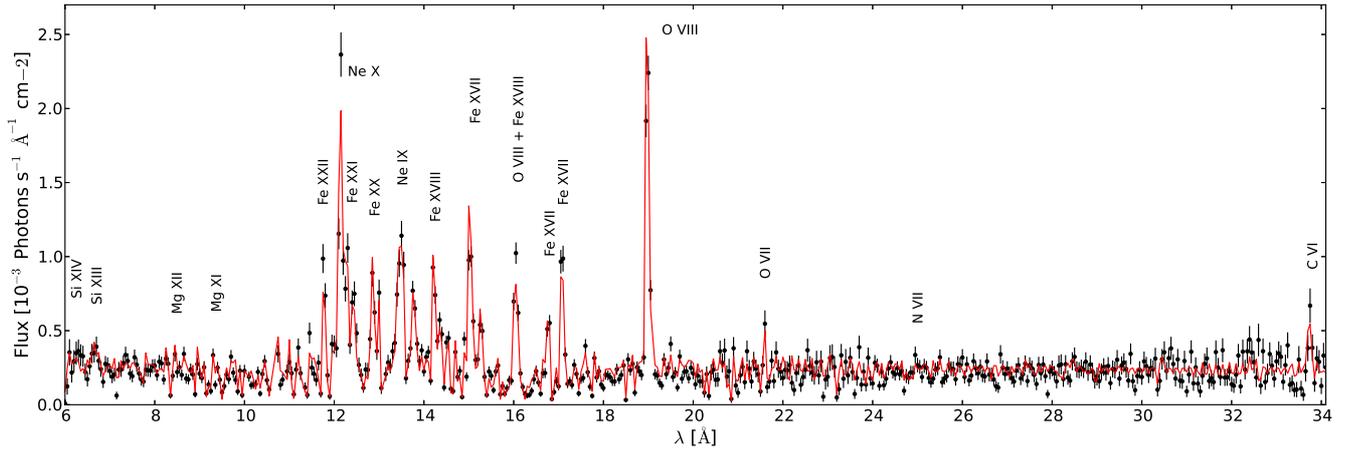}
 \caption{RGS spectrum (RGS1 and RGS2 combined) of Obs. ID 060317201. Plotted model is a simultaneous fit to all instruments (see EPIC spectra in Figure~\ref{fg:EPIC_spectra}.}
 \label{fg:RGS_spectrum}
\end{figure*}

Sample spectra and best-fitted model are presented in this section.
The X-ray spectra obtained from our first observation ID 060317201, as measured by all {\it XMM-Newton} instruments used in the analysis are shown in Fig.~\ref{fg:RGS_spectrum}. 
This observation has the lowest count rates (see Fig.~\ref{fg:pn_lc}).
Consequently, the spectra of the three other observations have slightly better signal-to-noise ratios, but are overall similar to those shown in Fig.~\ref{fg:RGS_spectrum}.
In Section~\ref{sec:Rotation_Effects}, each observation is divided into four segments, and the four iso-phase segments are then stacked, resulting in spectra that are, again, with numbers of photons and signal-to-noise ratios that are comparable to those of the individual observations. 
 
 \begin{figure*}[h]
 \centering
 \includegraphics[width=\textwidth]{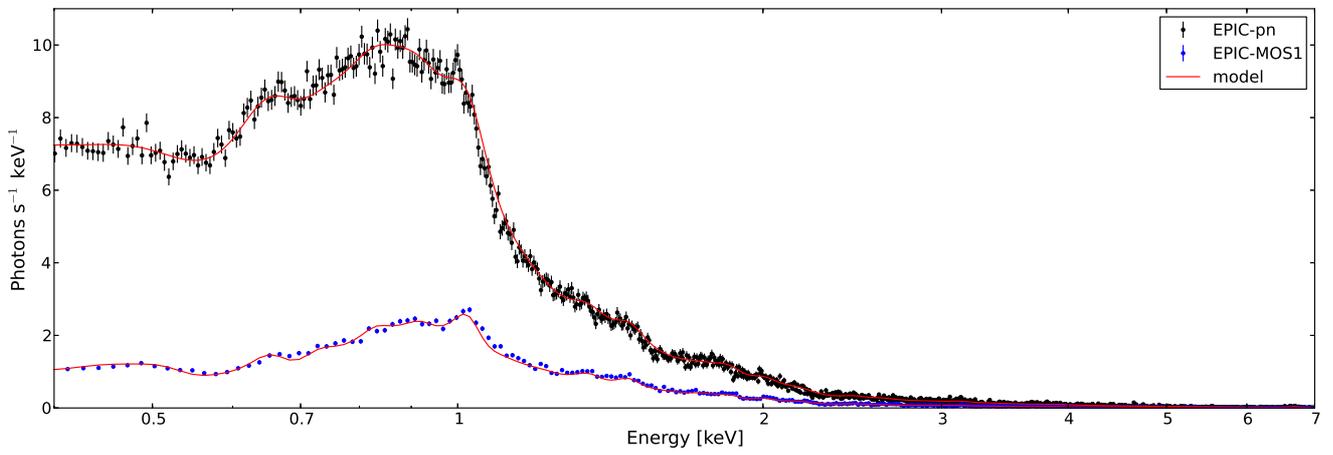}
 \caption{EPIC-pn and EPIC-MOS1 spectra of Obs. ID 060317201. Plotted model is a simultaneous fit to all instruments (see RGS spectrum in Figure~\ref{fg:RGS_spectrum}.}
 \label{fg:EPIC_spectra}
\end{figure*}

 \section{$IEM$ and $<kT>$ Error Estimates}
 \label{app:MC_simulation}
The $EMD$ solutions to coronal emission tend to be highly degenerate as emission measure can be traded between different, but close temperatures with little effect on the resulting spectrum. Thus, when representing the $EMD$ with isothermal components, the uncertainties in the normalizations of the various components are highly correlated, especially when many components are used. The correlation tends to be negative and increases as the components are closer in temperature, i.e., adjacent temperatures can compensate for each other.  This results in large errors on each individual temperature component, which can be seen in Fig.~\ref{fg:EMD}.
On the other hand, the exchange of emission measure between close-temperature components has only a small effect on the cumulative, or integrated emission measure ($IEM$ ). Still, the spectral line emissivities are temperature dependent and thus the $IEM$ is not perfectly conserved. In addition, there are the uncorrelated uncertainties associated with the limited statistics of the data. In order to estimate the highly nonlinear contributions of these effects on the uncertainties in the $IEM$, we allude a Monte-Carlo simulation.
 
The fitting algorithm in XSPEC gives, along with the best fit model, the covariance matrix of second derivatives calculated at the best-fit solution.
Here, we use this covariance matrix to redraw a random, new set of model $EMD$ values on the given temperature grid  (see Fig.~\ref{fg:IEM}) that is close in its statistical properties to the original best-fit model, in the sense that it produces spectral fits to the data that are almost as good as the best-fit model.
In order to generate a random set of numbers corresponding to a correlation matrix $R$, one can decompose $R$ into $R = U^{\top}U$.
If $X$ is a vector of uncorrelated random values, $X_c = XU$ is a set of correlated random values corresponding to $R$.
The process of generating random $EMD$s is repeated numerous times, and for each $EMD$ solution the corresponding $IEM(kT)$ is calculated at each temperature point on the grid.
Thus, we generate a large sample of $IEM(kT)$ values at each temperature point, from which we adopt the central 68\% as the equivalent of the 1-$\sigma$ uncertainty interval that is, then, plotted in Figures~\ref{fg:IEM} and \ref{fg:quarters_IEM}.
For each generated random $EMD$ we also calculate the emission-measure weighted mean temperature $<kT>$. The central 68\% of the $<kT>$ distribution is adopted as the 1-$\sigma$ uncertainty interval on $<kT>$, plotted in Figures~\ref{fg:kT_effective} and \ref{fg:quarters_effective_kT}.

\end{appendix}

\end{document}